\documentclass[12pt]{iopart}
\usepackage{iopams}
\expandafter\let\csname equation*\endcsname\relax
\expandafter\let\csname endequation*\endcsname\relax
\usepackage{amsmath}
\usepackage{subfig}
\usepackage{epsfig}
\usepackage{color}
\usepackage{graphicx}
\usepackage{graphicx,epstopdf}
\usepackage{hyperref}
\usepackage{cite}
\usepackage{commath}
\usepackage{color}
\usepackage{times}
\usepackage{graphicx}
\usepackage{graphicx,epstopdf}
\usepackage{dcolumn}
\usepackage{bm}
\usepackage{mathrsfs}
\usepackage{soul}
\usepackage{amsfonts}
\usepackage{float}

\usepackage{mathtools}
\usepackage{blkarray, bigstrut}
\usepackage{hyperref}

\usepackage{mathtools}

\usepackage{graphicx}
\usepackage{dcolumn}
\usepackage{bm}
\usepackage{graphicx}
\usepackage{xcolor}
\graphicspath{{./figures/}}
\usepackage{hyperref}
\usepackage{amsfonts}
\usepackage{amssymb}
\usepackage{amsmath}
\usepackage{notoccite}
\usepackage{lineno}
\usepackage[utf8x]{inputenc}
\usepackage{lineno}
\usepackage[T1]{fontenc}
\usepackage{mathptmx}
\usepackage[inline]{enumitem}

\begin{document}
	\title[Extreme events in globally coupled chaotic maps]{Extreme events in globally coupled chaotic maps}
	
	\author{ Sayantan Nag Chowdhury$^1$, Arnob Ray$^1$, Arindam Mishra$^2$, and Dibakar Ghosh$^1$}
	\address{$^1$Physics and Applied Mathematics Unit, Indian Statistical Institute, 203 B. T. Road, Kolkata-700108, India\\
		$^2$Division of Dynamics, Technical University of Lodz, Stefanowskiego 1/15, 90-924 Lodz, Poland}
	\ead{dibakar@isical.ac.in (D. Ghosh)}

	\date{\today}

\begin{abstract}
	
	
	Understanding and predicting uncertain things are the central themes of scientific evolution. Human beings revolve around these fears of uncertainties concerning various aspects like a global pandemic, health, finances, to name but a few. Dealing with this unavoidable part of life is far tougher due to the chaotic nature of these unpredictable activities. In the present article, we consider a global network of identical chaotic maps, which splits into two different clusters, despite the interaction between all nodes are uniform. The stability analysis of the spatially homogeneous chaotic solutions provides a critical coupling strength, before which we anticipate such partial synchronization. The distance between these two chaotic synchronized populations often deviates more than eight times of standard deviation from its long-term average. The probability density function of these highly deviated values fits well with the Generalized Extreme Value distribution. Meanwhile, the distribution of recurrence time intervals between extreme events resembles the Weibull distribution.  The existing literature helps us to characterize such events as extreme events using the significant height. These extremely high fluctuations are less frequent in terms of their occurrence. We determine numerically a range of coupling strength for these extremely large but recurrent events. On-off intermittency is the responsible mechanism underlying the formation of such extreme events. Besides understanding the generation of such extreme events and their statistical signature, we furnish forecasting these events using the powerful deep learning algorithms of an artificial recurrent neural network. This Long Short-Term Memory (LSTM) can offer handy one-step forecasting of these chaotic intermittent bursts. We also ensure the robustness of this forecasting model with two hundred hidden cells in each LSTM layer.
	
\end{abstract}

\maketitle

%

\section{Introduction} \label{introduction}

\par Human's spontaneous endeavor for understanding the unknown drives innovation by surpassing the limitations. The growing fascination to predict uncertainty benefits us to construct flexible tools for generating feasible explanations behind those puzzles. Despite substantial progress of such forecasting schemes, accurate prediction of uncertain future remains a grand challenge. And this challenge accelerates to few times in case of understanding and predicting extreme events. Extreme events \cite{mcphillips2018defining,ghil2011extreme}, in general, occur far off from the mean state (central tendency) of the skewed distribution \cite{frolov2019statistical}. Consequently, extreme events appear in the tail of the distribution with comparatively less frequency. Thus, the scientific community has too often dealt with a small number of data extracting information about the unpleasant long-lasting outcomes of extreme events. Recently, dynamical systems \cite{lucarini2016extremes,nag2021ee} are being recognized as a valuable contemporary theory to overcome the curse of limited data.

\par {The dynamical systems perspective facilitates propagating a large amount of data by evolving the equations of motions of the system forward in time. The temporal evolution of a suitable observable exhibiting extremely large or small excursions from the long-term average has attracted considerable attention among interdisciplinary studies. These recurrent excursions may resemble extreme events, as their irregular occurrence in time is seemingly random. Motivated by these analogies, many investigations take place, shifting the focus to the single and coupled dynamical systems.} Recently, Ray et al.\ \cite{ray2020extreme} demonstrated the emergence of extreme events under the accumulated effect of distributed damping parameter and repulsive interaction in a globally coupled network of heterogeneous Josephson junctions. Earlier, the underlying generating mechanisms of extreme events in excitable systems of FitzHugh–Nagumo units with different coupling topologies, including global and small-world network, are investigated thoroughly in Refs.\ \cite{ansmann2013extreme,karnatak2014route,ansmann2016self}. Occurrence of the dragon-king-like extreme events due to occasional irregular unison firing in two coupled Hindmarsh-Rose bursting neurons is inspected in Ref.\cite{mishra2018dragon}. The role of edges in the origination of extreme events is scrutinized in Refs.\ \cite{brohl2020identifying,kumar2020extreme}. {However, the study of extreme events has been less explored in coupled maps.} 

\par The statistical properties of extreme events in deterministic dynamical systems are analyzed in the seminal work \cite{nicolis2006extreme}. Ray et al.\ \cite{ray2019intermittent} successfully applied a threshold-activated coupling \cite{sinha1993adaptive} to suppress extreme events in two coupled Ikeda maps. The sudden relatively rare and significantly large explosive growth of population densities is recently reported in a network of coupled chaotic Ricker maps \cite{moitra2019emergence}. {Except for these few pioneering works, most of the earlier studies on extreme events deal with only continuous dynamical systems \cite{mishra2020routes,ray2020understanding,suresh2020parametric,sudharsan2021emergence,kaviya2020influence,bonatto2017extreme} to the best of our knowledge.} In this article, our results reveal the emergence of extreme events in globally coupled chaotic maps due to on-off intermittency \cite{platt1993off}. Cavalcante et al.\ \cite{cavalcante2013predictability} previously reported this mechanism of causing extreme events. On-off intermittency is also identified as the principal process for the genesis of extreme events in temporal networks of mobile agents \cite{9170822,chowdhury2019extreme} under the influence of attractive-repulsive co-existing interactions \cite{majhi2020perspective,chowdhury2020effect,mishra2015chimeralike,chowdhury2021antiphase}.

\par Quite astonishingly, the coupled maps have fascinating, realistic applications, including stock markets \cite{ma2004stochastic}, fluid dynamics \cite{aref1983integrable}, quantum field theories \cite{beck2002spatio}, optical systems \cite{perez1992nonstatistical}, and many more. Coupled map lattices \cite{kaneko2015globally,xie1996coherent,jalan2005synchronized,amritkar2005synchronized,kaneko1989chaotic,jost2001spectral} can offer a plethora of emerging dynamical behaviors, ranging from complete synchronization \cite{chowdhury2019convergence,rakshit2018synchronization,chowdhury2019synchronization}, splay state \cite{singha2016spatial}, chimera \cite{khaleghi2019chimera,parastesh2020chimeras} to clustering \cite{balmforth1999synchronized,popovych2002role} and spatiotemporal intermittency \cite{kaneko1992overview} to name but a few. We identify extreme events in the globally coupled maps near the synchronization manifold and utilize the benefits of the machine learning approach to predict 
the dynamics incorporating extreme events. Recently, model-free prediction of chaotic dynamics has become one of the mushrooming avenues of research that offers considerable benefits in forecasting extreme events without knowing the explicit equations of the model. Reservoir computing, one of those effective tools, can be applied to indicate early warning signals of extreme events successfully, as recently highlighted by Pyragas et al.\ \cite{pyragas2020using}. The performance of the neural networks for predicting chaotic dynamics of the two-dimensional H\'{e}non map is investigated rigorously in Ref.\ \cite{lellep2020using}. Several other data-driven methods enjoy widespread appreciation in the current literatures \cite{ray2021optimized,qi2020using,narhi2018machine,sapsis2018new,wan2018data,amil2019machine,refId0} for forecasting extreme events. Our study involves a special kind of Recurrent neural networks (RNN) \cite{abiodun2018state}, viz.\ Long Short-Term Memory (LSTM) \cite{schmidhuber1997long,vlachas2018data,qin2019comparison,guth2019machine,vlachas2020backpropagation,sangiorgio2020robustness,wang2021hybrid,yanan2020chaotic} and unveils its excellent flexibility to capture the dynamics of upcoming extreme events optimally.

\par Elman RNNs \cite{elman1990finding} require repeated multiplication of the gradient with the weight matrix during backpropagation through time \cite{werbos1988generalization}. This leads to the vanishing or exploding gradients problem depending on the recurrent weight matrix's negative or positive spectral radius. Besides, a vanilla RNN cannot remember long-term dependencies of data over a long time. The limitations of RNNs fuel the discoveries of LSTM \cite{schmidhuber1997long}, which can yield one-step-ahead of prediction, disregarding the problems of speed and stability in RNNs. Apart from multiple practical applications ranging from short-term traffic prediction \cite{zhao2017lstm} to handwriting \cite{graves2008novel} and speech \cite{li2015constructing} recognition, LSTM is recently found to be fruitful in detecting extreme events at Uber data \cite{laptev2017time} and flood forecasting \cite{le2019application}. In what follows, we are interested in exploring the emergence of extreme events in globally coupled chaotic maps. The necessary details of the model and the mechanism behind the formation of extreme events are the contents of Sec.\ \ref{Mechanism}. Section \ref{Statistics} is devoted to the statistical studies of extreme events. We discuss the range of coupling strength to provide a clear separation between the non-extreme and extreme events in Sec.\ \ref{Range}. We then proceed with the accurate one-step-ahead prediction of extreme events in Sec.\ \ref{Prediction}, whereas we summarize and conclude with a discussion of our findings in the last Sec.\ \ref{Conclusion}.



\section{Mechanism involved in the formation of extreme events} \label{Mechanism}

\par  The coupling through the mean-field in an ensemble of identical coupled oscillators can provide fascinating outcomes like explosive death \cite{dixit2021dynamic} and synchronous state \cite{doi:10.1063/5.0039879}. We consider the following equations

\begin{equation}
\begin{split}
x_{n+1}(i)=f(x_n(i))+\epsilon \big(\bar{f_n}-f(x_n(i)) \big),~ i=1,2,\cdots, N
\end{split}
\label{eq1}
\end{equation}

describing the dynamics of $N (\geq 2)$ globally coupled one-dimensional maps \cite{kaneko1990globally}. Here, $x_n(i)$ reflects the one-dimensional real state vector of the $i$-th map at the $n$-th iteration. The dynamics of each map is given by the $C^1$ smooth one-dimensional nonlinear function $f:\mathbb{R} \to \mathbb{R}$. $\bar{f_n}=\frac{1}{N} \sum_{j=1}^{N} f(x_n(j))$ depicts the arithmetic mean. The coupling parameter $\epsilon$ lies within the compact interval $[0,1]$. However, $\epsilon=0$ describes $N$ isolated maps. On the other hand, each map evolves synchronously after the first iteration for $\epsilon=1$. To determine the range of $\epsilon$ for the global stability of the synchronous state, we define the following function for any pair of $k$-th and $l$-th maps at the $n$-th iteration as,

\begin{equation}
\begin{split}
V_n(kl)=\big(x_n(k)-x_n(l) \big)^2.
\end{split}
\label{eq2}
\end{equation}

\par Clearly, this function is non-negative and it will vanish only when the $k$-th and $l$-th maps evolve to the same motion at the $n$-th iteration. Thus, if we can show that this function is monotonically decreasing, then it will converge to the greatest lower bound (infimum), as the function \eqref{eq2} is bounded below by $0$. The criteria of asymptotic global stability, hence, reduces to $V_{n+1}(kl) < V_n(kl)$. For large $N$, one can easily derive that the system \eqref{eq1} asymptotically synchronizes \cite{jost2001spectral} if

\begin{equation}
\begin{split}
1-\frac{1}{A} < \epsilon < 1+\frac{1}{A},
\end{split}
\label{eq3}
\end{equation}

where $A=sup \abs{f^{'}} \neq 0$ represents the least upper bound (supremum) of $f^{'}(x)$. We further perform linear stability analysis for the complete synchronization state. The Jacobian matrix is $\boldmath{J}=\boldmath{J_{0}} f^{'}$, where $f^{'}$ is the derivative at the synchronous state and $\boldmath{J_{0}}=\text{circ} \big(1-\epsilon+\frac{\epsilon}{N}, \frac{\epsilon}{N}, \frac{\epsilon}{N}, \cdots, \frac{\epsilon}{N}\big)$ is an $N \times N$ circulant matrix. The eigen values of $\boldmath{J_{0}}$ are $\mu_1=1$ and $\mu_j=1-\epsilon$ for $j=2,3,\cdots,N$. The eigenvector $\psi_1$ corresponding to the eigen value $\mu_1$ represents the synchronization manifold and hence, the amplification of the perturbation inside the one-dimensional invariant manifold $\{(x(1),x(2),\cdots,x(N))| x(1)=x(2)=\cdots=x(N)\}$ along this eigenvector does not destroy the coherence. The other $N-1$ identical eigen values help in determining the stability of the coherent state, as their corresponding eigen vectors $\psi_j$ for $j=2,3,\cdots,N$ evolve transverse to the synchronization manifold. With the help of the Lyapunov exponent $\lambda_1$ of the uncoupled map, the linear stablity analysis provides the critical condition \cite{balmforth1999synchronized} as follows,

\begin{equation}
\begin{split}
\epsilon_c = 1-\exp(-\lambda_1).
\end{split}
\label{eq4}
\end{equation}

Clearly, the synchronized state is stable for $\epsilon > \epsilon_c$ after a sufficient number of iterations (transient). In order to verify our findings, we choose the classical logistic map $f(x)=rx(1-x)$, where $r \in [0,4]$ is the nonlinear parameter. The globally coupled map \eqref{eq1} settles down to the fixed point $0$ for $r \in [0,1)$, and converges to the fixed point $1-\dfrac{1}{r}$ for $r \in (1,3)$ independent of the choice of the coupling strength $\epsilon$. For the numerical simulations, we set $f(x)$ in the chaotic regime with $r=4$ and consider $N=200$. As $\epsilon$ lies within $[0,1]$, the specific choice of the logistic map with $r=4$ leads to the global synchronization for $\epsilon \in \bigg(\dfrac{3}{4},1\bigg]$ as given in Eqn.\ \eqref{eq3}. On the other hand, $\lambda_1$ is a function of $r$. For $r=4$, the Lyapunov exponent of the single logistic map is $\lambda_1=\ln2$. Hence, the critical coupling strength $\epsilon_c$ becomes $\dfrac{1}{2}$ as per the relation \eqref{eq4}.

\par
\begin{figure*}[!t]
	\centerline{\includegraphics[width=0.95\textwidth]{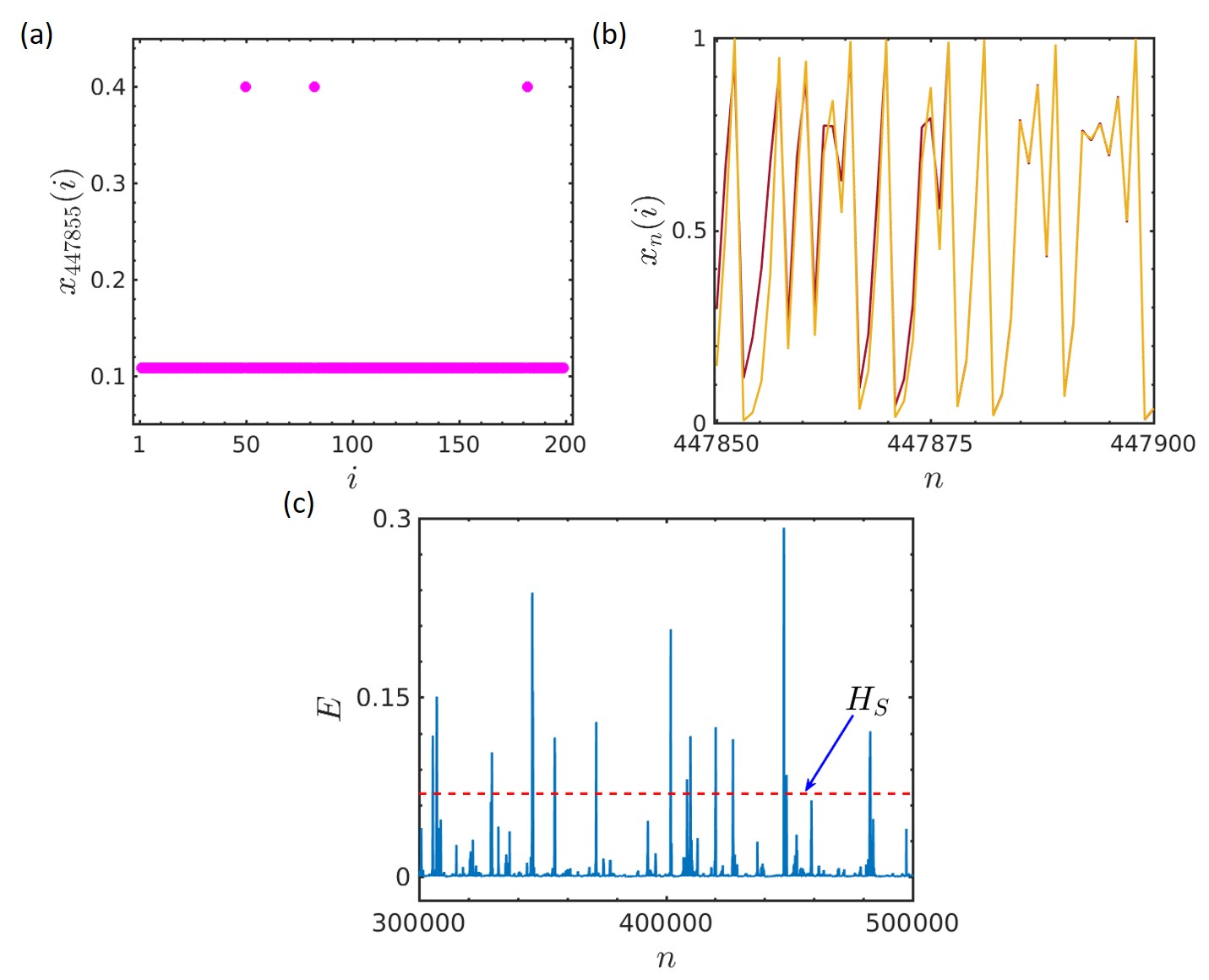}}
	\caption{(Color Online) (a) {\bf Advent of two synchronized clusters in the globally coupled maps}: We choose the coupling strength $\epsilon=0.4995$ slightly lesser than the critical coupling strength $\epsilon_c=0.5$. This $\epsilon_c$ is necessary for achieving complete synchronization of the system \eqref{eq1} with $N=200$ as per Eqn.\ \eqref{eq4}. The system \eqref{eq1} settles down to a partial synchronized state. The number of elements in each cluster varies depending on the initial conditions. (b) {\bf Time evolution of each logistic map}: We use the identical logistic maps $f(x)=rx(1-x)$ as the local dynamics with $r=4$. Each map reflects a chaotic behavior. Although those oscillators occasionally collapse to a single trajectory, mostly they possess a two-cluster state, as observed in the snapshot at a specific iteration $447855$ in subfigure (a). (c) {\bf On-off intermittency of the observable $E$}: We select two different maps, viz.\  $1$-st and $50$-th oscillators from two distinct clusters using subfigure (a) and calculate the Euclidean distance $E$ between them.  The temporal evolution of the observable $E$ contemplates chaotic intermittent bursts which cross the significant height $H_S=\langle E_n \rangle + 8\sigma$ from time to time.  $\langle E_n \rangle$ is the sample mean of the local maxima $E_n$ of $E$ gathered over sufficiently long $10^8$ iterations after the initial transient. $\sigma$ is the standard deviation of the whole simulated data. We will assess an event $E_n$ as an extreme event if it crosses the red horizontal dashed line $H_S$. This figure is drawn for a particular set of randomly distributed initial conditions within the interval $[0,1]$. We consider $3 \times 10^5$ iterations as the transient. For further details, please see the text.}
	\label{Figure1}
\end{figure*}

 We plot Fig.\ \eqref{Figure1} for $\epsilon=0.4995$ such that the difference between chosen $\epsilon$ and $\epsilon_c$ remains very small. All figures of this article, including this one, are plotted for a fixed set of initial conditions randomly distributed in the uniform interval $[0,1]$. All necessary fundamental codes along with this set of initial conditions utilized in this article are available at \cite{web_1}.  
 However, we verify that the results will remain unchanged for any other random choice of initial conditions for each map within the interval $[0,1]$. The choice of this $\epsilon=0.4995$ will not allow these $N=200$ coupled maps to settle down to a complete synchronized state as expected through our stability analysis. Figure \eqref{Figure1} (a) reveals the partial synchronization \cite{popovych2002role,panchuk2002stability} state of the system, where the entire system \eqref{eq1} splits into two groups. Particularly for the chosen set of initial conditions, one group contains exactly $N_1=3$ elements, and the other one possesses $N_2=N-N_1=197$ members. These numbers $N_1$ and $N_2$ depend on the choice of random initial conditions. All figures are inspected after $3 \times 10^5$ iterations. We plot the dynamics of each logistic map in Fig.\ \eqref{Figure1} (b), which demonstrates the temporal chaotic behavior of each cluster. Careful scrutiny of this subfigure unfolds the complex behavior of these two clusters. The distance between these two groups is not same in the entire time domain.

\par The temporal dynamics of two clusters \cite{popovych2002role} can be described as

\begin{equation}
\begin{split}
y_{n+1}(1)=(1-\epsilon)f(y_{n}(1))+\epsilon p_1 f(y_{n}(1))+\epsilon p_2 f(y_{n}(2)),~ \text{and}\\
y_{n+1}(2)=(1-\epsilon)f(y_{n}(2))+\epsilon p_1 f(y_{n}(1))+\epsilon p_2 f(y_{n}(2)) \hspace{0.85cm}
\end{split}
\label{eq5}
\end{equation}

where, $p_j=\dfrac{N_j}{N}$ for $j=1$ and $2$ with $N_1+N_2=N$. Here, without any loss of generality, we assume that the dynamical elements $y_{n}(1)$ and $y_{n}(2)$ satisfy the following relations of the synchronized oscillators after a viable relabelling as,

\begin{equation}
\begin{split}
x_{n}(1)=x_{n}(2)=\cdots=x_{n}({N_1})=y_{n}(1), ~\text{and}\\
x_{n}({N_{1}+1})=x_{n}({N_{1}+2})=\cdots=x_{n}({N})=y_{n}(2).
\end{split}
\label{eq6}
\end{equation}

When any one of $p_1$ or $p_2$ proceeds to the limit $0$, then the other one tends to $1$. For instance, if $p_1 \to 0+$, then $p_2 \to 1-$. Thus for $p_1=0$, Eqns.\ \eqref{eq5} govern the following equations,

\begin{equation}
\begin{split}
y_{n+1}(1)=(1-\epsilon)f(y_{n}(1))+\epsilon f(y_{n}(2)), ~\text{and}\\
y_{n+1}(2)=f(y_{n}(2)). \hspace{4.1cm}
\end{split}
\label{eq7}
\end{equation}

\par Noticeably, $y(2)$ does not depend on the variable $y(1)$ for $p_1=0$, although $y(1)$ depends on both the variables $y(1)$ and $y(2)$. Thus, these equations in the asymmetric cluster limit possess the skew product structure \cite{glendinning2003stability}.

\par We define a new variable $E=\abs{y(1)-y(2)}$ describing the Euclidean distance between two clusters. Figure \eqref{Figure1} (c) displays the temporal dynamics of $E$ after the initial transient. $E$ remains near zero (under the numerical accuracy) most of the iterations revealing the synchronization of the two clusters. However, this off-state ($E \approx 0$) is interrupted occasionally, leaving the trait of large amplitude random bursts. This non-zero value of $E$ is so high that it crosses the pre-defined threshold $H_S$ (shown in dashed red line in subfigure \eqref{Figure1} (c)) infrequently. This $H_S=\langle E_n \rangle + 8\sigma$ is treated here as the extreme event qualifying threshold \cite{chowdhury2019extreme}, where $E_n$ is the local maxima of $E$ after the initial transient of $3 \times 10^5$ iterations. We accumulate the values of $E$ over $10^8$ iterations. Then, we collect the values of $E_n$ excluding the transient data. $\langle E_n \rangle$ represents the sample mean of this gathered data of $E_n$, and $\sigma$ is the corresponding standard deviation. The choice behind this threshold $H_S$ is motivated by the fact that the extreme events qualifying threshold is generally taken as $\langle E_n \rangle + d\sigma$, where $d \in \mathbb{R}$ varies within the interval $[4,8]$ \cite{ray2020extreme,mishra2018dragon,9170822,chowdhury2019extreme,kaviya2020influence,mishra2020routes,suresh2020parametric,sudharsan2021emergence}. Evidently, we choose the upper bound $8$ of $d$, as it definitely portrays more relatively scattered events with strikingly higher amplitude and lower frequency. Moreover, Ref.\ \cite{chowdhury2019extreme} presented an analytical formulation for this particular choice of threshold, and designated this threshold as the significant height in analogy with the oceanic rogue waves \cite{ryszard1996ocean}. We employ this significant height $H_S$ borrowing the idea from Ref.\ \cite{chowdhury2019extreme} and whenever $E_n$ crosses $H_S$, we recognize it as an extreme event.

\par We rigorously investigate that these sporadic bursts of $E$ are not due to the appearance of floating nodes \cite{jalan2005synchronized}. Once a node joins a synchronized cluster, it will stay there forever and will never move to the other cluster. This intermittent bursts are completely due to the varying distance of the two clusters. We choose two distinct maps from two different clusters and study the temporal evolution of their Euclidean distance $E$ in Fig.\ \ref{Figure1} (c). For further evaluation of extreme events, we investigate their statistical properties in the following section.

\begin{figure}[!t]
	\centerline{\includegraphics[width=0.85\textwidth]{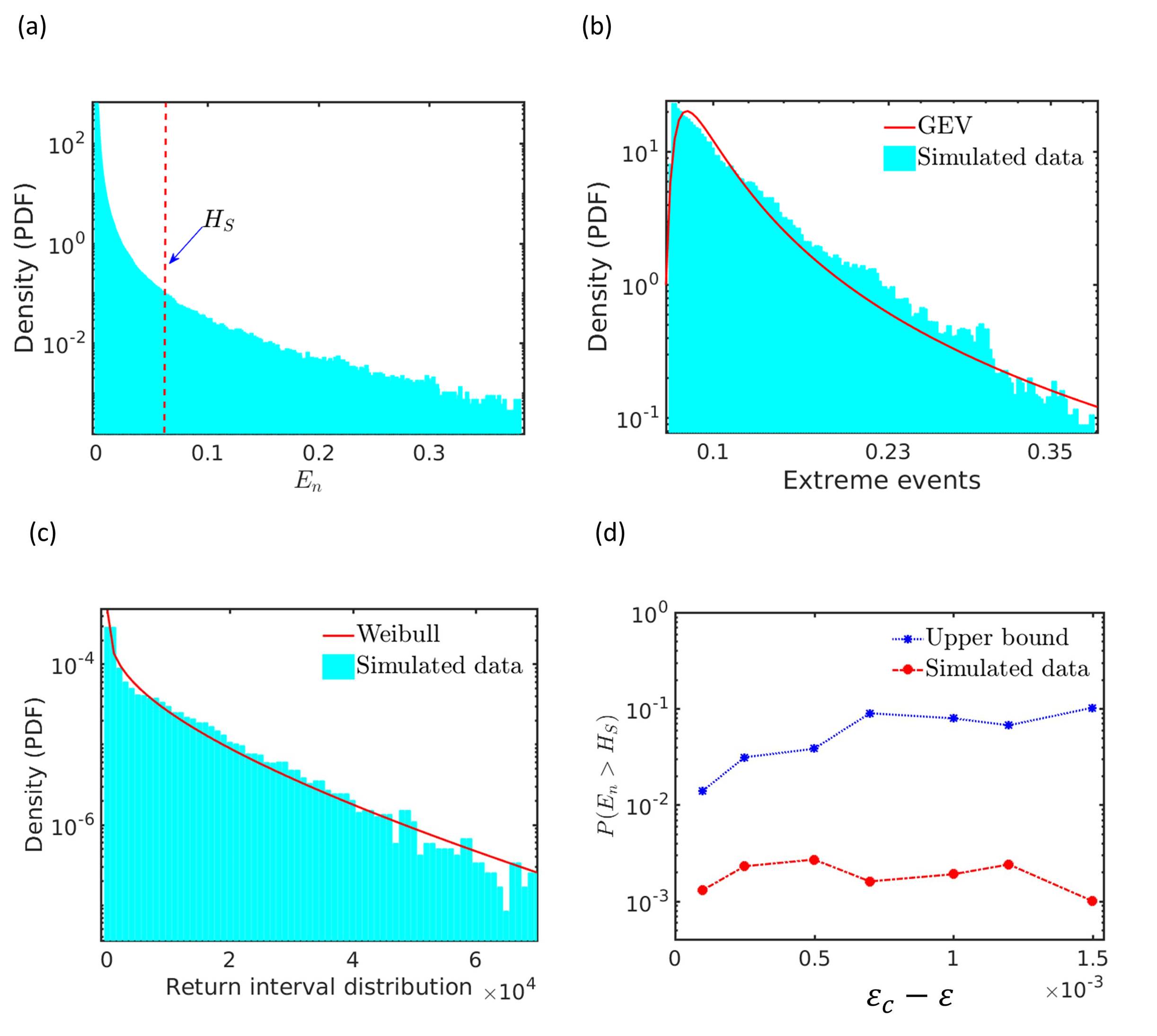}}
	\caption{(Color Online) (a) {\bf Non-normal distribution of the local maxima}: We accumulate local maxima $E_n>0$ of $E$ over sufficiently long $10^8$ iterations after repudiating the first $3 \times 10^5$ transients. We calculate $H_S$ over these data and draw this to identify the separation between non-extreme and extreme events. (b) {\bf Probability density function of extreme events}: Generalized extreme value distribution fits well with the data from the right-hand side of the significant height $H_S$ in subfigure (a). We use log likelihood to estimate the three parameters of the family given in Table \eqref{GEV}. (c) {\bf The distribution of return intervals for the extreme events within each burst}: We numerically investigate the successive manifestation of extreme events within each burst. For numerical simulation, once we detect an extreme event, we throw away subsequent extreme events (if it exists) from next $100$ iterations. This data is well fitted with Weibull distribution. The estimated scale parameter is $a=6782.39$ and the estimated shape parameter is $b=0.721932$. We also use the popular P-P (probability-probability) plot (not shown here) to confirm our claim. The subfigures (a-c) are drawn with $\epsilon=0.4995$ for the $N=200$ globally coupled identical logistic maps $f(x)=4x(1-x)$. (d) {\bf The probability of appearance of extreme events}: We compute the probability of occurrence of extreme events for $7$ different values of $\epsilon$ using the relation \eqref{eq12} and successfully validate it by plotting the upper bound using the inequality \eqref{eq11}. Here, $\epsilon_c=0.5$. We use the same set of random initial conditions drawn from $[0,1]$ for all results.}
	\label{Figure2}
\end{figure}

\section{Statistical information of extreme events} \label{Statistics}

\par Generally, Gaussian (normal) distribution, being symmetrically distributed about the mean, indicates undoubtedly that the data near the central tendency are comparatively more frequent in occurrence than the data far from the expected value. We assemble the data of local maxima $E_n$ of $E$ over a considerable interval of $10^8$ iterations after leaving the initial transient of $3 \times 10^5$ iterations. We plot the probability density function (PDF) of $E_n$ in Fig.\ \ref{Figure2} (a). We choose the same set of initial conditions and coupling strength $\epsilon=0.4995$ as in Fig.\ \ref{Figure1}. The red dashed vertical line $H_S$ denotes the extreme events indicating threshold. The value of the threshold for this particular choice of initial conditions is $H_S \approx 0.0693$. The left-hand side of $H_S$ contains all non-extreme events, while the right-hand side encompasses all extreme events. Clearly, the PDF of $E_n$ follows the non-normal distribution as a fair portion of data deviates more than $8\sigma$ from the mean $\langle E_n \rangle$. In addition, we use the statistical Kolmogorov-Smirnov test (K-S test) by assuming our simulated data are identical with the Gaussian distribution. 
As anticipated, the K-S test rejects the null hypothesis at the default $5\%$ significant level. These findings validate the signature of extreme events employed in Refs.\ \cite{akhmediev2010editorial,pisarchik2011rogue,9170822,chowdhury2019extreme}. More or less, the common notion of definition of extreme events in these Refs.\ \cite{akhmediev2010editorial,pisarchik2011rogue,9170822,chowdhury2019extreme} divulges the following characteristics, viz.\ 
\begin{enumerate}
	\item these events are aperiodic, possessing amplitude higher than $H_S=\langle E_n \rangle + 8\sigma$, and
	\item these recurrent events occur less frequently, maintaining the non-Gaussian distribution.
\end{enumerate}

In this article, we also maintain these two important properties in case of defining extreme events. We also represent the PDF of the amplitude of the extreme events in Fig.\ \ref{Figure2} (b). We draw the subfigures (a-c) of Fig.\ \ref{Figure2} using the MATLAB Distribution Fitter App. Our numerically simulated data (shown in cyan color) in Fig.\ \ref{Figure2} (b) is the same data, lying in the right of the significant height $H_S$ in Fig.\ \ref{Figure2} (a). We utilize the Freedman–Diaconis rule \cite{freedman1981histogram} for choosing the width of the bins in this histogram. We find this data are well fitted with the PDF of the generalized extreme value (GEV) distribution \cite{de2006extreme} given by 

\begin{equation}
\begin{split}
G(x)=\dfrac{1}{\beta} \exp \bigg(-\bigg(1+\gamma \dfrac{x-\alpha}{\beta}\bigg)^{-\dfrac{1}{\gamma}}\bigg)\times\bigg(1+\gamma \dfrac{x-\alpha}{\beta}\bigg)^{-\dfrac{1}{\gamma}-1}
\end{split}
\label{eq8}
\end{equation}

for $\beta \neq 0$ and $1+\gamma \dfrac{x-\alpha}{\beta} > 0$. Here, $\alpha, \beta>0, \gamma \neq 0$ stand for the location, scale, and shape parameters, respectively. As GEV distribution is a combination of three families (Gumbel, Fr\'{e}chet and Weibull), we get Type-II extreme value distribution for $\gamma > 0$. In contrast, we obtain the Type-III case for $\gamma < 0$. For $\gamma=0$, we have

\begin{equation}
\begin{split}
G(x)=\dfrac{1}{\beta} \exp \bigg(-\exp \bigg(- \dfrac{x-\alpha}{\beta}\bigg)-\dfrac{x-\alpha}{\beta}\bigg)
\end{split}
\label{eq9}
\end{equation}

representing Type-I case. Table \eqref{GEV} contains all the necessary information about the estimated parametric values of $\alpha$, $\beta$, and $\gamma$ for Fig.\ \ref{Figure2} (b). GEV distribution is applicable in various fields \cite{de2006extreme}, including it is the only possible limiting distribution (if it exists) of an adequately normalized maximum of a sample of independent and identically distributed random variables. 

\begin{table}[h!]
	\centering
	\begin{tabular}{|c| c | c|} 
		\hline
		Parameter & Estimate & Standard Error \\ [1ex] 
		\hline\hline
		$\alpha$ & $0.0891089$ & $8.7959 \times 10^{-5}$ \\ [1ex] 
		\hline
		$\beta$ & $0.0212706$ & $8.88201 \times 10^{-5}$\\[1ex] 
		\hline
		$\gamma$ & $0.60792$ & $0.00466071$ \\[1ex]
		\hline
	\end{tabular}
	\caption{Estimated parameter values for GEV distribution}\label{GEV}
\end{table}



\par We also study the statistics of the return intervals between extreme events in Fig.\ \ref{Figure2} (c). Recurrence interval has a broad range of applications in hazard estimations like the interarrival packet times on internet traffic \cite{antoniou2002log}, earthquakes \cite{corral2004long}, floods \cite{task1996hydrology}, and many more. We select the extreme events from each burst and collect the return interval between two consecutive such extreme events. For numerical implementation, once we detect an extreme event, then we ignore the extreme events (if it exists) for the successive $100$ iterations. The sample standard deviation of the gathered data is $1.0490 \times 10^4$ and the sample mean is $8.3016 \times 10^3$. Thus, the coefficient of variation (CV) is $1.2636$. This dimensionless number being greater than unity reveals a high variance of the data. Moreover, the CV of an exponential distribution is precisely $1$, which is less than the CV of our return intervals. Hence, we conclude that the return interval distribution of extreme events does not follow the exponential distribution. Earlier studies \cite{santhanam2008return,bunde2005long,eichner2006extreme} suggest these return intervals are generally approximated to any one of the following distributions, viz.\ (i) exponentially distributed, (ii) stretched exponential, and (iii) Weibull distribution. Memoryless is a vital property of the exponential distribution of non-negative real numbers. In contrast, long-term memory usually gives rise to a stretched exponential distribution of the return intervals \cite{bunde2005long}. Recently, Weibull distribution is also found to be a good representative for the distribution of recurrence time intervals of long-range correlated data \cite{santhanam2008return}. The PDF plotted in the semi-log scale resembles of a product of a stretched exponential and a power law. This PDF in Fig.\ \ref{Figure2} (c) is well fitted by the Weibull distribution. Estimation of the scale and shape parameter is given in Table \eqref{Weibull}.

\begin{table}[h!]
	\centering
	\begin{tabular}{|c | c| c | c|} 
		\hline
		Parameter & Estimate & Standard Error \\ [1ex] 
		\hline\hline
		$a$ & $6782.39$ & $90.4573$ \\ [1ex] 
		\hline
		$b$ & $0.721932$ & $0.00522843$\\[1ex] 
		\hline
	\end{tabular}
	\caption{Estimated parameter values for Weibull distribution}\label{Weibull}
\end{table}

\par Here, $a$ is the scale parameter and $b$ is the shape parameter. The flexibility of the Weibull distribution allows it to model in an astonishingly diverse array of data sets, ranging from weather forecasting to reliability engineering and failure analysis. The PDF of the two parameter Weibull distribution is given by

\begin{equation}
\begin{split}
W(x)=\dfrac{b}{a} \big(\dfrac{x}{a} \big)^{b-1} {e}^{-\big(\dfrac{x}{a}\big)^{b}}.
\end{split}
\label{eq10}
\end{equation}

\par Although $E$ remains zero for a large number of iterations as illustrated through Fig.\ \ref{Figure1} (c), $E_n$, being the local maxima of $E$, is positive-valued observable. Evidently, $H_S=\langle E_n \rangle + 8\sigma$ is also positive quantity and varies with each different choice of initial conditions, even for fixed $\epsilon$ and $N$. We come up with an upper bound for the probability of occurrence of extreme events using Markov's inequality \cite{cohen2015markov} as follows

\begin{equation}
\begin{split}
P(E_n > H_S) \leq \dfrac{\langle E_n \rangle}{H_S} < 1.
\end{split}
\label{eq11}
\end{equation}

\par We compute the probability of occurrence of extreme events over the interval $[0.4985,0.4999]$ of the coupling strength $\epsilon$ with the same set of initial conditions used in Fig.\ \ref{Figure1}. To calculate its probability, we iterate the system \ref{eq1} with $N=200$ identical logistic maps for $10^8$ iterations and then discard the initial $3  \times 10^5$ data treating them as transient. Then, we calculate the number of local maxima $E_n$ of $E$ over that sufficiently long data. After that, we simulate the significant height $H_S$ in each case and count the number of extreme events, i.e., $E_n$ which cross $H_S$. Hence, we apply the classical definition of probability as follows

\begin{equation}
\begin{split}
\text{Probability of occurrence of extreme events}=\dfrac{\text{Number of extreme events}}{\text{Number of }E_n}.
\end{split}
\label{eq12}
\end{equation}

\par We plot these values for $\epsilon_c-\epsilon$ in Fig.\ \ref{Figure2} (d) along with the upper bound using the relation \ref{eq11}. Here, $\epsilon_c=0.5$ is the required critical coupling strength for the emergence of complete synchronization as calculated through the Eqn.\ \eqref{eq4}. One should notice that the upper bound proposed in the inequality \ref{eq11} is not, by any means, the least upper bound. Even though we cannot observe any clear trend between the numerically calculated probabilities, one can anticipate that as the distance between $\epsilon_c-\epsilon$ increases, the chaotic bursts of $E$ become more intermittent due to the instability of synchronization invariant manifold $\{(x(1),x(2),\cdots,x(N))| x(1)=x(2)=\cdots=x(N)\}$. Consequently, these sporadic bursts become more frequent and become less likely to cross the threshold $H_S$. We will discuss this result in detail in the next section.

\section{Range of coupling strength for the emergence of extreme events} \label{Range}

\begin{figure*}[!t]
	\centerline{\includegraphics[width=1.0\textwidth]{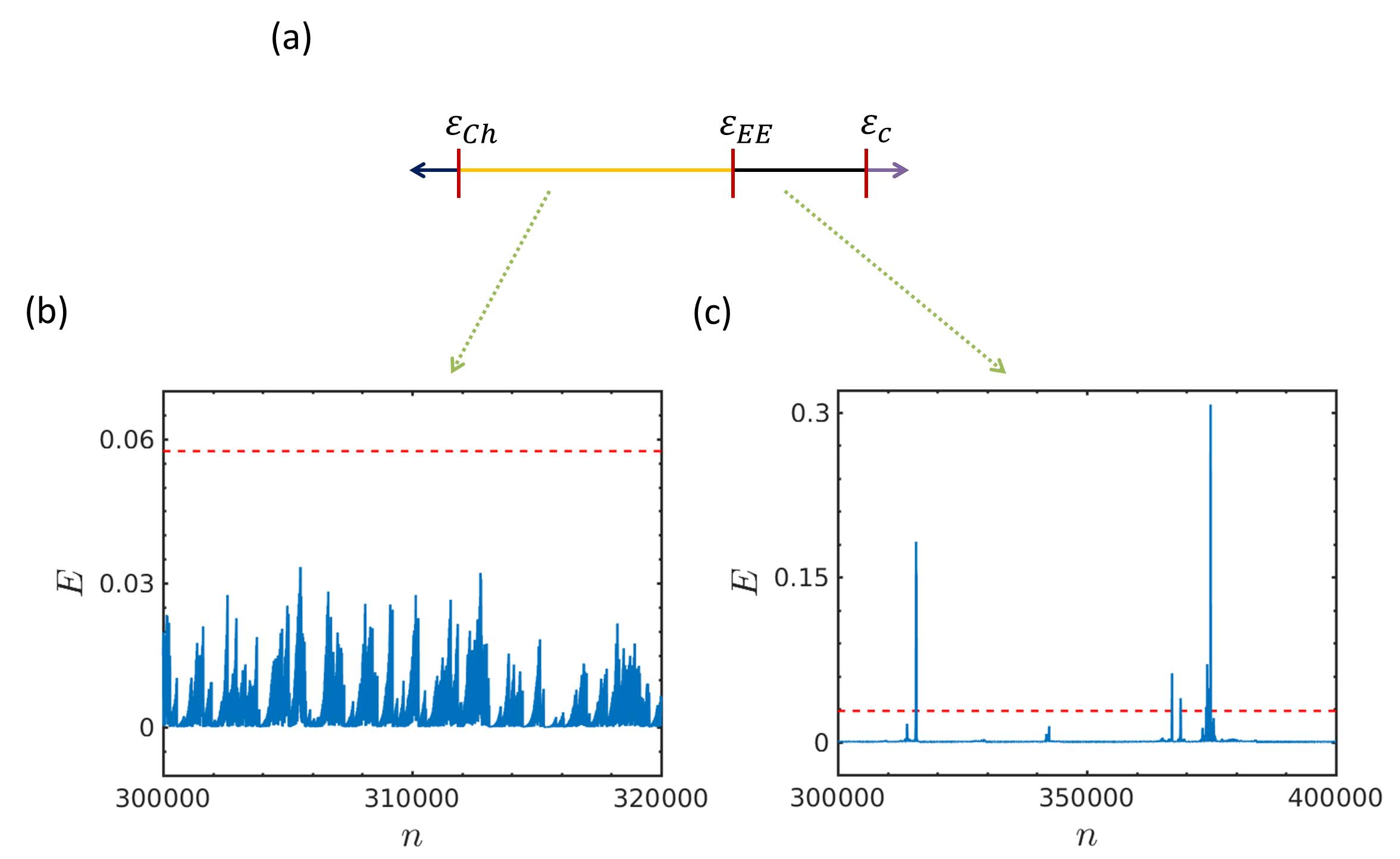}}
	\caption{(Color Online) {\bf The range of coupling strength separating the non-extreme and extreme events}: We split an interval of $\epsilon$ over four distinct groups. Dark blue interval of $\epsilon < \epsilon_{ch}$ exhibits periodic two  clusters, whereas the remaining interval contains a two-cluster chaotic attractor for $(\epsilon_{ch}, \epsilon_{c})$. The orange interval $(\epsilon_{ch}, \epsilon_{EE})$ displays the interval encompassing non-extreme events. We come up with the subfigure (b) for $\epsilon=0.496$, where the red dashed horizontal line is the significant height $H_S=\langle E_n \rangle + 8\sigma$. Evidently, $E$ flunks to cross this threshold $H_S$. Nevertheless $E$ crosses $H_S$ intermittently for $\epsilon=0.4999$ in subfigure (c). The black interval $(\epsilon_{EE},\epsilon_{c})$ of subfigure (a) incorporates the interval of extreme events. The purple interval of $\epsilon > \epsilon_c=0.5$ presents the coherent chaotic attractor. For simulation, we consider $N=200$ globally coupled maps with identical logistic maps $f(x)=4x(1-x)$ as the local dynamics. For further details, please see the text.}
	\label{Figure3}
\end{figure*}

\par We contemplate the dynamical behavior of the system \eqref{eq1} of $N=200$ identical logistic maps in Fig.\ \eqref{Figure3} with the same initial conditions as used in all previous figures. This inspection will help us to determine the range of $\epsilon$ corresponding to extreme events. Figure \eqref{Figure3} (a) reflects a one-dimensional parameter range of the coupling strength $\epsilon$, where we identify numerically four distinct behaviors. For $\epsilon < \epsilon_{ch} \approx 0.43303012$ (dark blue interval of Fig.\ \eqref{Figure3} (a)), the system settles down to a two cluster state. The behavior of each map is periodic in time, and accordingly, the distance between two clusters $E$ is also periodic. Hence, we never expect the manifestation of extreme events in this range. The system falls in a chaotic two cluster state within the interval $(\epsilon_{ch}, \epsilon_{c})$. The purple regime of $\epsilon > \epsilon_{c}$ in Fig.\ \eqref{Figure3} (a) gives rise to a coherent chaotic state, as anticipated from our stability analysis. However, we fail to provide an analytical estimation to decide the range of $\epsilon$ exhibiting extreme events. That's why we examine its range through numerics. Our numerical study suggests extreme events arise only within the interval $(\epsilon_{EE},\epsilon_{c})$, where $\epsilon_{EE} \approx 0.4961745$. The interval $(\epsilon_{ch},\epsilon_{EE})$ yields the two cluster chaotic attractor, but their Euclidean distance $E$ becomes too frequent to cross $H_S$. To reveal this nature, we choose $\epsilon=0.496$ from the non-extreme (orange) range of Fig.\ \eqref{Figure3} (a) and select the $1$-st and $2$-nd maps from two different clusters. The difference between these two maps is plotted in Fig.\ \eqref{Figure3} (b). Clearly, the intermittent random bursts of $E$ do not cross the extreme event qualifying height $H_S$ (red dashed horizontal line) in this figure. Similarly, we set $\epsilon = 0.4999$ from the interval (black) of occurrence of extreme events in Fig.\ \eqref{Figure3} (a). The system again falls in a partially synchronized state, and we choose two different maps, viz.\  $98$-th and $99$-th maps corresponding to two distinct clusters. The distance between these two maps occasionally crosses the significant height $H_S$ in Fig.\ \eqref{Figure3} (c). The aperiodic random bursts attest to the emergence of extreme events.

\par One should emphasize that although we plot the open interval $(\epsilon_{ch}, \epsilon_{c})$ continuously by the black solid line in Fig.\ \eqref{Figure3} (a). However, one may find a set of Lebesgue measure zero within this interval, which contains few points of coupling strength that unable to produce extreme events. One such point is $\epsilon=0.49775$. For this choice of $\epsilon$, we accumulate sufficiently long data of $E_n$ over $10^8$ iterations after discarding the initial transient $3 \times 10^5$. We observe that these long iterates of $E_n$ fail to cross the derived $H_S$. We draw this Fig.\ \eqref{Figure3} with the same set of initial conditions as used in Fig.\ \eqref{Figure2}. However, we verify the results remain unchanged if we choose other random initial conditions of each element
from the uniform interval $[0,1]$. Although, other initial conditions may slightly change the value of $\epsilon_{ch}$ and $\epsilon_{EE}$ .

\begin{figure*}[!t]
	\centerline{\includegraphics[width=0.90\textwidth]{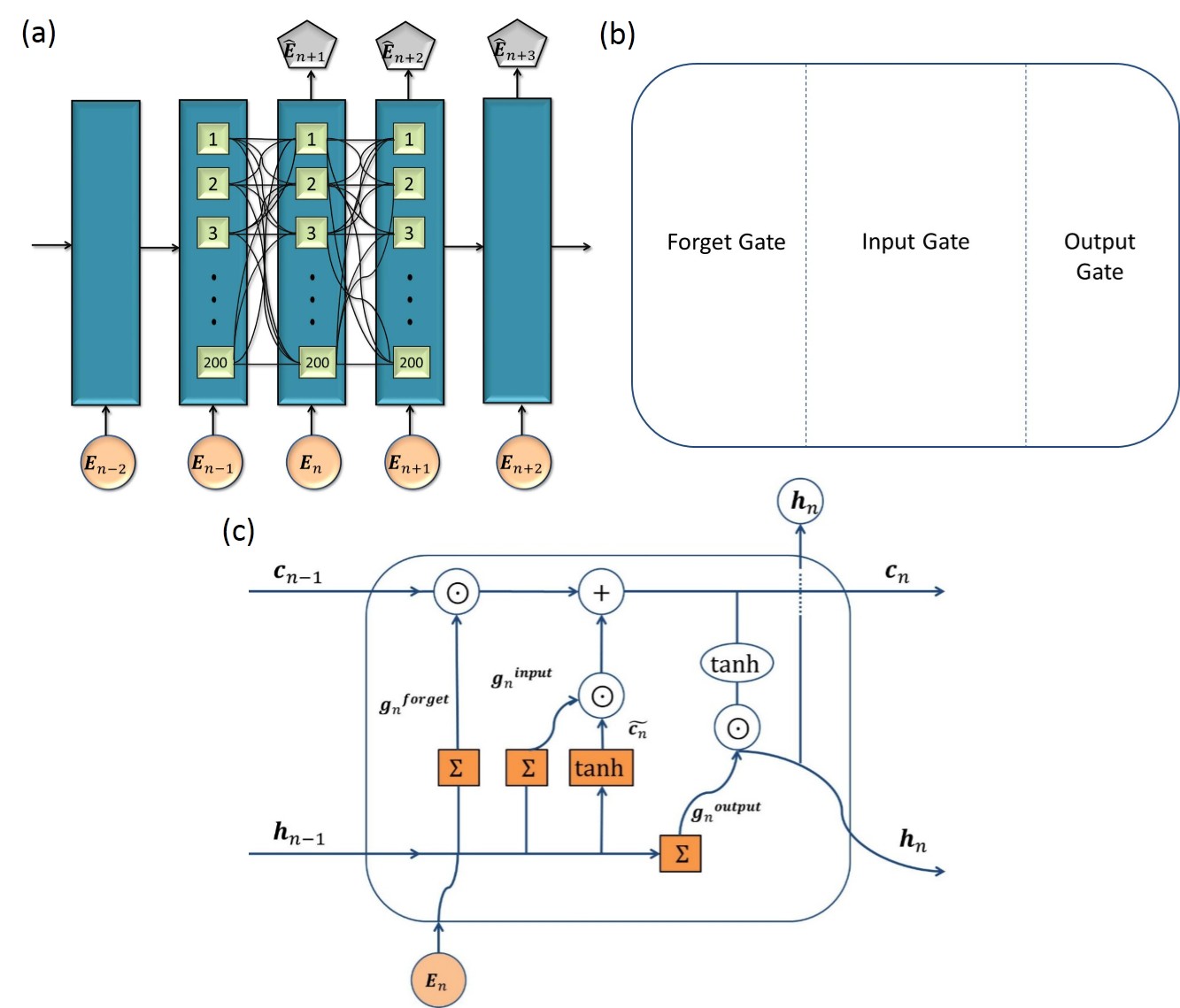}}
	\caption{(Color Online) {\bf Architecture of Long Short Term Memory (LSTM)}: In subfigure (a), we present a detailed schematic design of an artificial recurrent neural network (RNN), which can avoid long-term dependency problems. Each layer contains $200$ LSTM neurons globally connected to all the LSTM neurons of the next layer. For clear visualization, we show all LSTM neurons explicitly only in three layers. We train the RNNs with previous $(n-1)$ data ($90\%$ of the accumulated data). Each LSTM cell consists of three different gates, as shown in subfigure (b). The forget gate decides whether the information from the cell state is required for the LSTM or not. The input gate helps in learning new relevant information from the input to the cell state. This gate also includes a sigmoid function like the forget gate. The output gate passes the updated filtered version of information from the current $n$-th iteration to the following $(n+1)$-th iteration.  Each LSTM has to deal with three things, viz.\ (i) the hidden state $\mathbf{h}_{n-1}$, i.e., the output of the last cell, (ii) the input $\mathbf{E}_{n}$, i.e., the new information available at the current time step, and (iii) the accessible information $\mathbf{c}_{n-1}$ of the previous cell after the previous time step. Subfigure (c) displays an LSTM cell, where ellipses and circles stand for point-wise operations, while rectangles indicate neural network layer operations. We unveil a concise description of this diagram accompanying all the related details in the main text.}
	\label{Figure4}
\end{figure*}

\par Lastly, we want to conclude this section with the fact that this appearance of extreme events is a generic signature in globally coupled chaotic maps, and our choice of the identical one-dimesional logistic map as the chaotic local map is only a paradigmatic example for illustrating the phenomena. 
 The intermittent transition \cite{balmforth1999synchronized,kaneko1989chaotic,kaneko1992overview} from the synchronized state to a two cluster attractor always allows a range of coupling strength for the origination of extreme events near the synchronization manifold in the system \eqref{eq1} consisting of one-dimensional chaotic maps.

\section{Short-term prediction of extreme events} \label{Prediction}

\par After the statistical characterization of extreme events, we focus on their prediction utilizing Long Short-Term Memory, a deep learning architecture. The recurrent neural nets, viz.\ long short-term memory (LSTM) networks used in this article is found to be capable of one-step-ahead forecasting of chaotic time series. {LSTM cell can process data sequentially and keeps its hidden state through time. Using the univariate time series, we create the regression data set \{${\bf X}$, ${\bf Y}$\} by taking a lag of $1$ (for example, if $X_n$ be the actual time iteration, then $Y_n=X_{n+1}$, $n=1,2,\cdots$, sample size). And using the training data set \{${\bf X}$, ${\bf Y}$\}, we build LSTM net so that we can deal with the vanishing gradient problem that can be encountered when training traditional RNNs. Thus, it is important to note that the dimension of the feature vector is one. So, sequence input with one dimension is passed in LSTM with $200$ hidden units in a  fully connected layer. Here, the loss function used to compute the weight matrix connecting different layers is a mean-squared error.} We iterate the system \eqref{eq1} with $\epsilon=0.4995$ for $10^8$ iterations and collect the distance $E$ between the two clusters after deleting first $3 \times 10^5$ transient. Then we calculate the local maxima $E_n$ of this whole gathered data of $E$. We use the first $165000$ data points of $E_n$ consisting of few extreme events, out of which $90\%$ data are used for the training. The remaining $10\%$ data points are treated as a testing set. The schematic sketch of neural architectures is shown in Fig.\ \eqref{Figure4} (a), where each layer comprises $200$ nodes for the LSTM. This number of hidden cells in each LSTM layer is determined using the cross-validation technique \cite{goodfellow2016deep}. $\widehat{\mathbf{E}_{n+1}}=F_{h}^{o}(\mathbf{h}_n)$ is the prediction for the observable $\mathbf{E}_{n+1}$ depending on the input observable $\mathbf{E}_n \in {\mathbb{R}}^{d_o}$. $F_{h}^{o}$ is the hidden to output mapping. $d_o$ and $d_h$ are the dimension of the input and hidden states respectively. This variant RNN consists of two internal states (i) cell state $\mathbf{c}_n \in {\mathbb{R}}^{d_h}$, and (ii) hidden state $\mathbf{h}_n=F_{h}^{h}(\mathbf{E}_n,\mathbf{h}_{n-1}) \in {\mathbb{R}}^{d_h}$, respectively. $F_{h}^{h}$ is the hidden to hidden mapping, which is given by the following equations

\begin{equation}
\begin{split}
\mathbf{g}_{n}^{\hspace{0.2cm}\mathbf{forget}}=\Sigma_{forget}\big( \mathbf{W_{forget}}\big[\mathbf{h}_{n-1},\mathbf{E}_n\big]+\mathbf{{\xi}_{forget}} \big),\\
\mathbf{g}_{n}^{\hspace{0.2cm}\mathbf{input}}=\Sigma_{input}\big( \mathbf{W_{input}}\big[\mathbf{h}_{n-1},\mathbf{E}_n\big]+\mathbf{{\xi}_{input}} \big),\\
\widetilde{\mathbf{c}_n}=\tanh\big( \mathbf{W_{cell}}\big[\mathbf{h}_{n-1},\mathbf{E}_n\big]+\mathbf{{\xi}_{cell}} \big),\\
\mathbf{c}_n=\mathbf{g}_{n}^{\hspace{0.2cm}\mathbf{forget}} \odot \mathbf{c}_{n-1}+ \mathbf{g}_{n}^{\hspace{0.2cm}\mathbf{input}} \odot \widetilde{\mathbf{c}_n},\\
\mathbf{g}_{n}^{\hspace{0.2cm}\mathbf{output}}=\Sigma_{hidden}\big( \mathbf{W_{hidden}}\big[\mathbf{h}_{n-1},\mathbf{E}_n\big]+\mathbf{{\xi}_{hidden}} \big), \hspace{0.2cm} \text{and}\\
\mathbf{h}_n=\mathbf{g}_{n}^{\hspace{0.2cm}\mathbf{output}} \odot \tanh \big(\mathbf{c}_{n}\big), \hspace{0.2cm} \text{with} \hspace{0.2cm} \mathbf{c}_0=\mathbf{0} \hspace{0.2cm} \text{and} \hspace{0.2cm} \mathbf{h}_0=\mathbf{0}.
\end{split}
\label{eq13}
\end{equation}

\begin{figure*}[!t]
	\centerline{\includegraphics[width=0.90\textwidth]{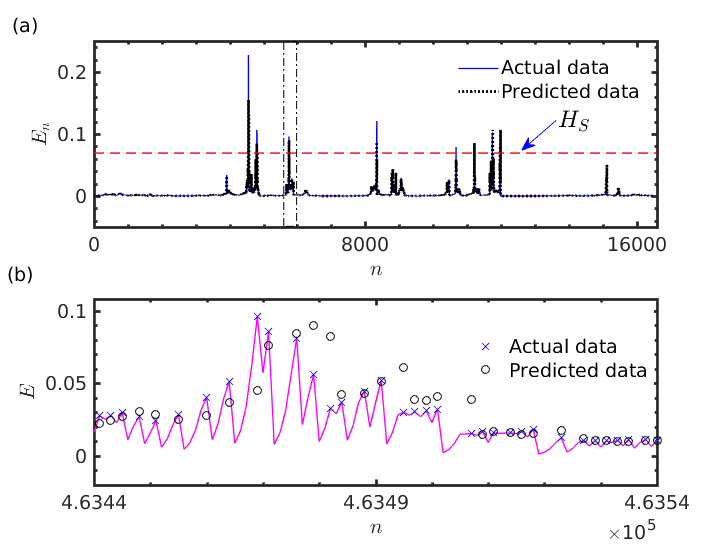}}
	\caption{(Color Online) {\bf Prediction of the local maxima $E_n$ of $E$}: (a) We plot the observed values (solid blue) of $E_n$ and corresponding predicted values (dotted black) simultaneously during the testing phase in the upper panel. The red dashed line $H_S$ is the extreme events qualifying threshold. Here, a short horizon of the test set containing $16499$ data points is used to validate our result. This figure attests that the prediction of the random chaotic bursts of $E_n$ consisting of few extreme events using a deep learning framework is successfully done with $mse=0.000011589$ and $nse=0.7907$. (b) For better visualization, an enlarged version of a portion (dashed black box) of the subfigure (a) is drawn in the lower panel. Here, a segment of time iterations (magenta) is displayed, where values of local maxima $E_n$ (blue cross) are termed as actual data. We perform the one-step-ahead prediction and denote the predicted values corresponding to $E_n$ by the black circle. This framework is also able to forecast the amplitude of extreme events besides non-extreme events. Here, non-extreme events depict those events with $E_n \leq H_S$. The initial conditions are the same as Fig.\ \eqref{Figure1}, and the actual data are simulated using the system \eqref{eq1} with $N=200$ identical logistic maps $f(x)=4x(1-x)$. }
	\label{Figure5}
\end{figure*}

\par Here, $\odot$ is the element-wise product (Hadamard product). Each LSTM is composed of three gates to protect and control the cell state, which carries the sequence input initially. $\mathbf{g}_{n}^{\hspace{0.2cm}\mathbf{forget}},\mathbf{g}_{n}^{\hspace{0.2cm}\mathbf{input}},\mathbf{g}_{n}^{\hspace{0.2cm}\mathbf{output}} \in {\mathbb{R}}^{d_h}$ represent the activation vectors of respective gates. $\mathbf{{\xi}_{forget}},\mathbf{{\xi}_{input}},\mathbf{{\xi}_{cell}}$ and $\mathbf{{\xi}_{hidden}}$ are the bias vectors. $\Sigma_{forget},\Sigma_{input},\Sigma_{hidden}$ are the sigmoid functions. The weight matrices $\mathbf{W_{forget}},\mathbf{W_{input}},\mathbf{W_{cell}},\mathbf{W_{hidden}} \in {\mathbb{R}}^{d_h \times (d_h+d_o)}$ along with the bias vectors need to be learned during training. The forget gate decides how much information should be appended in the final product of the cell state at each iteration $n$. This final product, i.e., the hidden state, is treated as the input of the next layer. The average of these outcomes across all hidden states is the final predicted output. The second last relation of Eqn.\ \eqref{eq13} suggest that depending on $\Sigma_{hidden}$, we get an output gate's activation vector $\mathbf{g}_{n}^{\hspace{0.2cm}\mathbf{output}}$. Hyperbolic tangent ($\tanh$) function acts on the cell state generating a value between $-1$ and $1$. The last relation of Eqn.\ \eqref{eq13} will produce the hidden state, which will give the prediction $\widehat{\mathbf{E}_{n+1}}$. This forecast $\widehat{\mathbf{E}_{n+1}}$ is a linear function $F_{h}^{o}$ given by the following relation

\begin{equation}
\begin{split}
\widehat{\mathbf{E}_{n+1}}=\mathbf{W_{o}}\mathbf{h}_n,
\end{split}
\label{eq14}
\end{equation}

where $\mathbf{W_{o}} \in {\mathbb{R}}^{d_o \times d_h}$. One such LSTM neuron is shown in Figs.\ \eqref{Figure4} (b-c). For a detailed description of LSTM networks, please see the pioneer work of Hochreiter et al.\ \cite{schmidhuber1997long}.

\par This LSTM network can give rise to the one-step-ahead future prediction depending on each time step of the input sequence as illustrated in Fig.\ \eqref{Figure4}. We plot the predicted output and the test data simultaneously for a short horizon in Fig.\ \eqref{Figure5} (a). To determine the accuracy of the convergence between these two data sets, we calculate the arithmetic mean of the squares of the Euclidean distance between $\mathbf{E}$ and $\widehat{\mathbf{E}}$ as follows

\begin{equation}
\begin{split}
mse\big(\mathbf{E},\widehat{\mathbf{E}}\big)=\dfrac{1}{l} \mathbf{e}^T \mathbf{e}=\dfrac{1}{l} \sum_{k=1}^{l} \big(e_i\big)^2,
\end{split}
\label{eq15}
\end{equation}

where $e_i$ is the difference between the forecasted outcomes and the actual value. $\mathbf{e}$ is the $l \times 1$ matrix and $l$ denotes the number of data points in the testing data. This loss function is a positive valued metric tending to zero for an accurate forecast. The $mse$ for Fig.\ \eqref{Figure5} (a) with  $l=16499$ is $0.000011589$. However, Ref.\ \cite{sangiorgio2020robustness} criticizes this mean squared error due to non-normalization for the variability of the data. Hence, we use a different measure as follows

\begin{equation}
\begin{split}
nse\big(\mathbf{E},\widehat{\mathbf{E}}\big)=1-\dfrac{mse\big(\mathbf{E},\widehat{\mathbf{E}}\big)}{mse\big(\mathbf{E},\langle \mathbf{E} \rangle\big)},
\end{split}
\label{eq16}
\end{equation}

where $\langle \mathbf{E} \rangle$ is the mean of the observed data. This measure is widely known as the Nash–Sutcliffe model efficiency coefficient being a normalized version of the $mse$, where $mse\big(\mathbf{E},\langle \mathbf{E} \rangle\big)$ is the data
variance. This metric $nse$ assumes the upper bound $1$ for the ideal scenario of exact forecasting. The $nse$ is $0.7907$ for Fig.\ \eqref{Figure5} (a) determining the predictive skill of our setup. The existing literature suggests an architecture promises more predictive performance for $nse \to 1-$. An enlarged portion of Fig.\ \eqref{Figure5} (a) is highlighted in Fig.\ \eqref{Figure5} (b).

\begin{figure}[!t]
	\centerline{\includegraphics[width=0.50\textwidth]{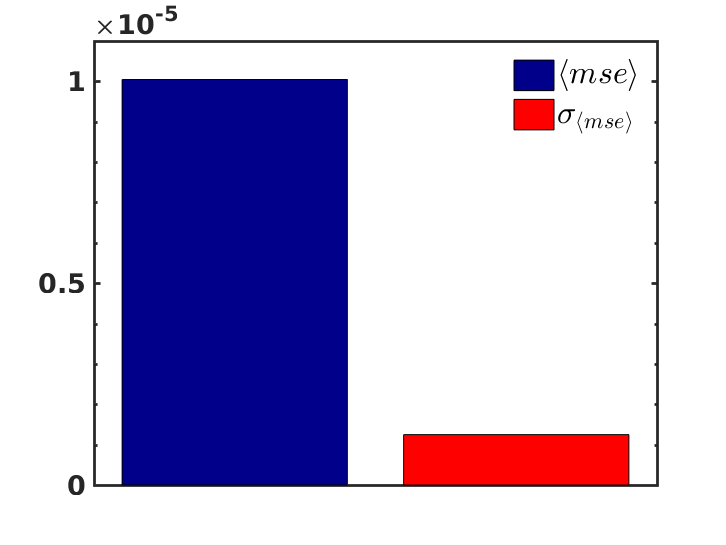}}
	\caption{(Color Online) {\bf Goodness of prediction and robustness of the forecasting model}: We perform $30$ independent trials of the prediction method and calculate the mean squared error (mse), a statistical measure to quantify the prediction accuracy of event forecasting. We plot mean ($\langle mse \rangle$) and fluctuation ($\sigma_{\langle mse \rangle}$) of mse over $30$ realizations in this figure. We evaluate $\langle mse \rangle=0.000010047$, and $\sigma_{\langle mse \rangle}=0.0000012476$. The small value of  $\langle mse \rangle$ signifies the good prediction as well as the small value of  $\sigma_{\langle mse \rangle}$ indicates that our prediction scheme is robust over the realizations.}	
	\label{Figure6}
\end{figure}

\par In fact, we contemplate the robustness of the used deep learning technique in Fig.\ \eqref{Figure6}. We define the following equations

\begin{equation}
\begin{split}
\langle mse \rangle=\dfrac{1}{h}\sum_{i=1}^{h} mse^i, \hspace{0.2cm} \text{and}\\
\langle nse \rangle=\dfrac{1}{h}\sum_{i=1}^{h} nse^i.\hspace{0.9cm}
\end{split}
\label{eq17}
\end{equation}

Here, $h$ is the number of trials. $mse^i$ and $nse^i$ are the respective $mse$ and $nse$ at the $i$-th realization. We take independent $h=30$ realizations and delineate the average $\langle mse \rangle$ with their corresponding standard deviation $\sigma_{\langle mse \rangle}$ in Fig.\ \eqref{Figure6}. $\langle nse \rangle$ for these $30$ independent trials is $0.81856$. It is evident that this $\langle nse \rangle$ is higher than the $nse$ of Fig.\ \eqref{Figure5}. Figures \eqref{Figure5} and \eqref{Figure6} reflect the successful one-step forecasting of the local maxima $E_n$ of $E$ with sufficiently moderate accuracy. These data sets possess few extreme events crossing the dashed line $H_S$ as shown in Fig.\ \eqref{Figure5}. We find our trained model agrees well with the observed chaotic data consisting of extreme events.

{
\section{Discussion and Conclusion} \label{Conclusion}

\par Recent decades have seen many efforts to understand the mechanisms behind the recurrent emergence of extreme events in dynamical systems and complex systems. Attainment of giant extraordinary values in a short time of a relevant observable is generally known as an extreme event. These recurrent short-lasting events deviate significantly from the long-term average of their statistical distribution. Due to its practical importance, this topic gains widespread recognition among interdisciplinary researchers. In spite of such leaps of progress, these investigations provide only a partial understanding of the mechanisms of extreme events and their predictions. The scientific community of nonlinear dynamics brings few successful signs of progress by considering investigations through models following differential equations. However, most of these studies are performed by contemplating continuous dynamical systems. Theoretical investigations on maps for an understanding of the dynamics of extreme events are not well explored yet. Only a few studies are reported on discrete dynamical systems \cite{nicolis2006extreme,ray2019intermittent,moitra2019emergence} along this line to the best of our knowledge. 

\par Moreover, most of the existing studies on globally coupled networks require repulsive interaction \cite{9170822,ray2020extreme} for the appearance of extreme events. The co-existence of attractive-repulsive coupling of suitable strength is necessary for the emergence of extreme events in globally coupled identical Stuart-Landau oscillators as per the numerical investigations of Ref.\ \cite{9170822}. Ray et al.\ \cite{ray2020extreme} also observed the signature of extreme events under repulsive mean-field interaction in a network of heterogeneous oscillators. Although, Ansmann et al.\ \cite{ansmann2013extreme} studied such ``self-generating and self-terminating" events under solely attractive coupling in globally diffusively coupled inhomogeneous FitzHugh–Nagumo units. But, parameter mismatch plays a crucial role in the formation of extreme events there.

}

\par We have investigated the emergence of extreme events in a globally coupled network of one-dimensional chaotic maps. We have calculated the critical coupling strength $\epsilon_c$, independent of the system size $N$, for achieving a synchronized chaotic state. Before this coupling strength, the system settles down to an asymmetric two-cluster attractor. The distance $E$ between these two clusters occasionally travels far away and possesses some large values.  These intermittent bursts of $E$ are so large that they infrequently deviate more than eight times the standard deviation from the mean of the local maxima $E_n$ of $E$ for a suitable range of coupling strength. These highly deviated values of $E$ are classified as extreme events if they cross the significant height $H_S$. The reason behind this specific choice of threshold is supplemented using the existing literature \cite{chowdhury2019extreme}. We have numerically determined the range of $\epsilon$ for the occurrence of extreme events in Sec.\ \eqref{Range}. Generalized extreme value distribution further corroborates the statistical signature of these events. We have evaluated the probability of the appearance of extreme events for some specific values of $\epsilon$ using Eqn.\ \eqref{eq12} and the procedure given in Sec.\ \eqref{Statistics}. For instance, this probability is $0.0027$ for the system \eqref{eq1} with $N=200$ coupled identical logistic maps $f(x)=4x(1-x)$ and $\epsilon=0.4995$. These results confirm that the appearance of extreme events is very fewer in frequency. The return time distribution of these extreme events resembles of Weibull distribution.

\par Understanding of extreme events brings to the limelight lots of open queries. The interdisciplinary research of complex systems plays a decisive role in providing valuable insights into these long-standing issues. We have contemplated the class of globally coupled one-dimensional chaotic maps hoping that our analysis may enhance the understanding of these non-equilibrium phenomena. This article has several folds with insightful results, including (i) the mechanisms triggering extreme events (see Sec.\ \eqref{Mechanism}), (ii) the statistical aspect of extreme events (see Sec.\ \eqref{Statistics}), and (iii) the prediction of extreme events (see Sec.\ \eqref{Prediction}). {We analyze these three essential themes in the studies of extreme events by placing the paradigmatic one-dimensional chaotic identical logistic maps in each node of the globally coupled network. The mechanisms of formation of extreme events may unravel in gaining insightful information regarding the onset of intermittency transition for globally coupled chaotic maps. Additionally, this route of two-state intermittent nature may unfold the possibility of mitigation of extreme events within chaotic dynamical systems. By tuning the coupling parameter, one can easily avoid such occasional giant events (see Sec.\ \eqref{Range}). And this mitigation strategy is only possible due to our precise knowledge of the mechanism viz.\ self-organized on-off intermittency triggering extreme events. In addition, 
	statistical studies are promising tools that help in drawing inferences regarding the frequency and probability of occurrence of extreme events from the samples \cite{sapsis2021statistics}. Nevertheless, knowledge of the mechanism behind the generation of extreme events may not always help in predicting extreme events. That’s why we take advantage of the machine learning approach to predict the dynamics incorporating extreme events. There are several examples of model-free predictions such as one-day-ahead forecasting electricity consumption \cite{devaine2013forecasting, nowotarski2016improving}, forecasting hourly solar irradiance \cite{mutavhatsindi2020forecasting}, predicting El Ni{\~n}o events with a lead time of one-year \cite{wang2021hybrid} to name a few. The deep learning strategies are capable of learning the essential chaotic dynamics directly from data and can deal with this grand undesirable challenge.} We are successfully able to predict the chaotic dynamics of $E_n$ consisting of extreme events with moderate accuracy. We hope that the perceived study may reveal a thorough understanding of this crucial phenomenon. The employed trained LSTM predictors have predictive power and practical applicability within the scope of viable forthcoming research and might help to comprehend the forecast of extreme events in realistic scenarios.

	\subsection*{Acknowledgment}
	
	S.N.C. is supported by the Council of Scientific \& Industrial Research  (CSIR)  under project No. 09/093(0194)/2020-EMR-I. {A.R. wants to thank Tanujit Chakraborty for their valuable discussions.} 

	\section*{DATA AVAILABILITY}
	
	Any data that support the findings of this study are included within the article. 
	

	\section*{References}
	\bibliographystyle{iopart-num}
	\bibliography{extreme}

\end{document}